\documentstyle[12pt]{article}

\font\titlefont=cmbx10 scaled \magstep3
\begin{document}
\input{epsf}

\begin{flushright}
TUTP-96-1  \\  gr-qc/9604052  \\ 
April 26, 1996
\vspace*{1cm}
\end{flushright}

\begin{center}
{\titlefont GRAVITONS AND LIGHTCONE \\
\vskip 0.2in
FLUCTUATIONS II: \\
\vskip 0.2in
CORRELATION FUNCTIONS }
\vskip .4in
L.H. Ford \\
\vskip .2in
Institute of Cosmology\\
Department of Physics and Astronomy\\
Tufts University\\
Medford, Massachusetts 02155\\
\vskip .3in
N.F. Svaiter \\ 
\vskip .2in
Centro Brasileiro de Pesquisas Fisicas-CBPF \\ 
Rua Dr. Xavier Sigaud 150\\ 
Rio de Janeiro, RJ 22290-180, Brazil \\
\end{center}

\vskip .3in

\begin{abstract}
A model of a fluctuating lightcone due to a bath of gravitons is further
investigated. The flight times of photons between a source and a detector
may be either longer or shorter than the light propagation time in the 
background classical spacetime, and will form a Gaussian distribution
centered around the classical flight time. However, a pair of photons
emitted in rapid succession will tend to have correlated flight times.
We derive and discuss a correlation function which describes this
effect. This enables us to understand more fully the operational
significance of a fluctuating lightcone. Our results may be combined with
observational data on pulsar timing to place some constraints on the 
quantum state of cosmological gravitons.  
\end{abstract}
\newpage

\baselineskip =14pt

\section{Introduction}

     In a previous paper \cite{F95}, henceforth I, the problem of lightcone 
fluctuations due to gravitons was discussed. A bath of gravitons in a squeezed
vacuum state, or a thermal state, was shown to produce fluctuations of the 
spacetime metric, which
in turn produce lightcone fluctuations. (A squeezed vacuum state is the state
in which relict gravitons from the early universe are expected to be found
\cite{Grishchuk}.) The propagation time of a classical light pulse \cite{pulse}
over a distance $r$ is no longer precisely $r$ 
\cite{units}, but undergoes fluctuations around a mean value of $r$.
In I, it was shown that the mean deviation from the classical propagation time
is 
\begin{equation}
\Delta t = \frac{\sqrt{\langle \sigma_1^2 \rangle}}{r} \,, \label{eq:Dt}
\end{equation}
where $\langle \sigma_1^2 \rangle$ is the mean square fluctuation in the 
geodesic interval function. Let $\sigma(x,x')$ be one-half of
the squared geodesic distance between a pair of 
points $x$ and $x'$. In the presence of a linearized metric
perturbation $h_{\mu\nu}$, 
\begin{equation}
\sigma = \sigma_0 + \sigma_1 + O(h_{\mu\nu}^2) \, .
\end{equation}
Here $\sigma_0 = \frac{1}{2} (x-x')^2$ is the flat space interval function,
and $\sigma_1$ is the first order shift in $\sigma$, which becomes a quantum
operator when the metric perturbations are quantized, The expectation values
of $\sigma_1^2$ are formally divergent, so $\langle \sigma_1^2 \rangle$ is
understood to be a renormalized expectation value, the difference between
the expectation value in a given state and in the Minkowski vacuum state. 
Note that we are assuming that the metric fluctuations are produced solely
by the bath of gravitons. More generally, quantum matter fields will experience
stress tensor fluctuations which will act as an additional source of metric
fluctuations \cite{F82,Kuo}. 

Equation~(\ref{eq:Dt}) arises from a calculation given in I of the expectation
value of the retarded Green's function in a squeezed vacuum state of 
gravitons, which yielded
\begin{equation}
\Bigl\langle G_{ret}(x,x') \Bigr\rangle = 
{{\theta(t-t')}\over {8\pi^2}} \sqrt{\pi \over {2\langle \sigma_1^2 \rangle}}
\; \exp\biggl(-{{\sigma_0^2}\over {2\langle \sigma_1^2 \rangle}}\biggr)\, .
                                           \label{eq:retav}
\end{equation}
 This form is valid for the case that $\langle \sigma_1^2 \rangle > 0$. 
Equation~(\ref{eq:retav}) reveals that the delta-function behavior of the
classical Green's function, $G_{ret}$, has been smeared out into a Gaussian
function which is peaked around the classical lightcone. This means that 
a light pulse is equally likely to traverse a distance in less than the 
classical propagation time as it is to traverse the interval in a longer time.

In I, explicit forms of $\langle \sigma_1^2 \rangle$ were given for particular
quantum states, including a single mode squeezed vacuum state and a thermal
state in the long wavelength limit. One of the purposes of the present paper
is to generalize these calculations, in particular to a bath of thermal 
gravitons in the short wavelength limit. This will be done in 
Sect.~\ref{sec:Bath}. 

The primary purpose of this paper will be to calculate and interpret 
the correlation function which relates the flight time variations of a 
pair of successive photons. In general, such a pair of photons may have
correlations which causes the expected difference in their flight times to be 
less than $\Delta t$. If one wishes to use observational data to search for
flight time variations due to lightcone fluctuations, 
it is essential to understand 
these correlations. In Sect.~\ref{sec:corrfnt}, a general formula for the
correlation function $\Bigl\langle G_{ret}(x_2,x_1) G_{ret}(x'_2,x'_1) 
\Bigr\rangle - \Bigl\langle G_{ret}(x_2,x_1) \Bigr\rangle
\Bigl\langle G_{ret}(x'_2,x'_1) \Bigr\rangle$ is obtained. In 
Sect.~\ref{sec:Flight}, this function is used to determine when pairs of
photons are correlated, and to calculate the mean variation in flight times,
$\delta t$. In Sect.~\ref{sec:Bounds}, we first review the effects of
classical gravity waves upon pulse arrival times, and the bounds which
pulsar timing data yield upon a background of such classical waves. We
then discuss the bounds which this data place upon gravitons in a squeezed
vacuum state. Our results are summarized and discussed in 
Sect.~\ref{sec:Summ}.

\section{Calculation of $\langle \sigma_1^2 \rangle$  for a 
Thermal Bath of Gravitons} \label{sec:Bath}

     In I, $\langle \sigma_1^2 \rangle$ was calculated for a thermal bath of
gravitons in the low temperature limit. In this section, we wish to generalize
this calculation to arbitrary temperature, and in particular obtain the high
temperature limit. In the transverse-tracefree gauge, the graviton two-point
function is expressible in terms of that for a massless scalar field: \cite{FP}
\begin{equation}
\langle h_{ij}(x)h^{ij}(x') \rangle = 2\langle \varphi(x) \varphi(x') \rangle 
   \, .
\end{equation}
From this expression, and Eqs. (49) and (61) of I, we obtain
\begin{equation}
\langle \sigma_1^2 \rangle =  {1\over {15}}\,r^2 \, 
\int_{0}^{r} dz \int_{0}^{r} dz'
\:\, \langle \varphi(x) \varphi(x') \rangle \, , \label{eq:sigma1ex}
\end{equation}
where the integrals are to be evaluated along the unperturbed null geodesic,
We take this geodesic to have a spatial extent of $r$ in the $z$ direction,
so $x = (z,0,0,z)$ and $x' = (z',0,0,z')$. Using the fact that the 
two-point function is a function of $\rho = |{\bf x} -{\bf x'}| =|z -z'|$, 
we may change variables of integration and write
\begin{equation}
\langle \sigma_1^2 \rangle =  {2\over {15}}r^2 
\int_{0}^{r} d\rho \, (r - \rho)
\:\, \langle \varphi(x) \varphi(x') \rangle \, . \label{eq:sigma1ex2}
\end{equation}

Note that the two-point function and the Hadamard function, $G(x,x')
=\frac{1}{2} \langle \{ \varphi(x) \varphi(x')\} \rangle$, are related
by
\begin{equation}
\langle \varphi(x) \varphi(x') \rangle = G(x,x') + 
\frac{i}{2} \Bigl[G_{adv}(x,x') - G_{ret}(x,x') \Bigr] \, ,
\end{equation}
where $G_{adv}(x,x')$ and $G_{ret}(x,x')$ are the advanced and retarded
Green's functions, respectively. Because the latter are proportional to
delta functions of the form $\delta(t-t' \pm |{\bf x} -{\bf x}'|)$, they will
not contribute to the renormalized thermal function, which is expressible as
an image sum over imaginary time. Thus for our purposes, the two-point and
Hadamard functions are identical. Some of the properties and limits of the
renormalized thermal two-point function, which we will denote by 
$G_{RT}(x,x')$, are discussed in the Appendix. 

A true thermal bath has both its characteristic wavelength, 
$\lambda_g = \beta$, 
and its graviton density determined by the temperature, $\beta^{-1}$. We
will also be interested in more general baths with an approximately
Planckian spectrum but an arbitrary amplitude. The mean squared 
amplitude of the metric fluctuations is characterized by the quantity
\begin{equation}
h^2 = \frac{1}{30} \langle h_{ij}(x)h^{ij}(x) \rangle \, . \label{eq:h2def}
\end{equation}
In the case of a thermal bath
\begin{equation}
h^2 = \frac{1}{180 \, \beta^2}\, , \label{eq:h2}
\end{equation}
so if we let
\begin{equation}
\langle \varphi(x) \varphi(x') \rangle = 180\, \beta^2\, h^2\, G_{RT}(x,x') \,,
                                                   \label{eq:2ptfnt}
\end{equation}
we obtain the two-point function for a more general bath with a Planckian
spectrum. In the following discussion, we will often use the symbols $\beta$
and $\lambda_g$ interchangeably. However, in general $\beta$ is understood to 
be the inverse temperature for a thermal bath, and $\lambda_g$ the mean
wavelength for a more general bath.

In the short wavelength (high temperature limit), we may use the asymptotic 
form for the thermal two-point function on the lightcone, 
Eq.~(\ref{eq:grthi_n}), to write
\begin{equation}
\langle \sigma_1^2 \rangle \approx  {3\over \pi}\, r^2\, \lambda_g h^2\,  
\int_{a\lambda_g}^{r} d\rho \, (r - \rho)
\:\, \frac{1}{\rho} \, . \label{eq:sigma1int}
\end{equation}
Recall that this asymptotic form is valid only for $\rho > a \lambda_g$, where
$a \gg 1$ is a constant. Thus we have introduced a cutoff at $\rho = a\lambda_g$
in the integral. This introduces an error of $O(\lambda_g^0)$, which is small
in the $\lambda_g \rightarrow 0$ limit. Evaluation of the integral in 
Eq.~(\ref{eq:sigma1int}) yields
\begin{equation}
\langle \sigma_1^2 \rangle \approx {3\over \pi}\, h^2\, r^3\, \lambda_g
 \,\biggl[\ln \biggl( \frac{r}{\lambda_g} \biggl) + c_1 \biggl] 
\, . \label{eq:sigma1gen}
\end{equation}
where $c_1$ is a constant of order unity.

   In I, the corresponding expression was derived for the low temperature 
(long wavelength) limit, where it was shown that
\begin{equation}
\langle \sigma_1^2 \rangle \approx  h^2\, r^4
\, , \label{eq:sigma1genlow} 
\end{equation}
for a general long wavelength graviton bath ($\lambda_g \gg r$). 
Note that comparison of Eqs.~(\ref{eq:sigma1gen}) and (\ref{eq:sigma1genlow})
reveals that $\langle \sigma_1^2 \rangle$ always grows with increasing 
$r$, but somewhat more slowly in the large distance limit where
$r > \lambda_g$.

\section{The Correlation Function}
\label{sec:corrfnt}

    In this section, we wish to derive and interpret an expression for the
correlation function
\begin{equation}
C(x_2,x_1;x'_2,x'_1) \equiv 
\Bigl\langle G_{ret}(x_2,x_1) G_{ret}(x'_2,x'_1) \Bigr\rangle 
- \Bigl\langle G_{ret}(x_2,x_1) \Bigr\rangle
\Bigl\langle G_{ret}(x'_2,x'_1) \Bigr\rangle \, .
\end{equation}
Our starting point is the Fourier representation of the retarded Green's
function,
\begin{equation}
G_{ret}(x_2,x_1) = {{\theta(t_2-t_1)}\over {8\pi^2}} \int_{-\infty}^{\infty}
                d\alpha\, e^{i\alpha \sigma_0}\, e^{i\alpha \sigma_1}\, .
                                                         \label{eq:gretrep}
\end{equation}
We may follow a procedure analogous to that used in I to obtain 
$\Bigl\langle G_{ret}(x_2,x_1) \Bigr\rangle$ to find
\begin{equation}
\Bigl\langle G_{ret}(x_2,x_1) G_{ret}(x'_2,x'_1) \Bigr\rangle =
   \frac{\theta(t_2-t_1) 
\theta(t'_2-t'_1)}{128 \pi^3\, \sqrt{B}}\, \exp \biggl(-\frac{A}{2B} \biggr)
 \, , \label{eq:Gsq}
\end{equation}
where
\begin{equation}
A = \sigma_0^2 \langle {\sigma'_1}^2 \rangle + 
     {\sigma'_0}^2 \langle \sigma_1^2 \rangle 
    -  2\sigma_0 \sigma'_0 \langle \sigma_1 \sigma'_1 \rangle \, ,
\end{equation}
and
\begin{equation}
B =  \langle \sigma_1^2 \rangle  \langle {\sigma'_1}^2 \rangle 
    -  \langle \sigma_1 \sigma'_1 \rangle^2 \, .
\end{equation}
Here $\sigma_0$ and $\sigma_1$ refer to the pair of points $(x_2,x_1)$, whereas
$\sigma'_0$ and $\sigma'_1$ refer to $(x'_2,x'_1)$.
Equation~(\ref{eq:Gsq}) is valid only if $ B >0$. 

    Let us consider the limit in which $\langle \sigma_1 \sigma'_1 \rangle$ is
small, so that
\begin{equation}
A \approx \sigma_0^2 \langle {\sigma'_1}^2 \rangle + 
     {\sigma'_0}^2 \langle \sigma_1^2 \rangle \, ,
\end{equation}
and
\begin{equation}
B \approx  \langle \sigma_1^2 \rangle  \langle {\sigma'_1}^2 \rangle 
    \, .
\end{equation}
In this limit, $\Bigl\langle G_{ret}(x_2,x_1) G_{ret}(x'_2,x'_1) \Bigr\rangle 
\approx \Bigl\langle G_{ret}(x_2,x_1) \Bigr\rangle
\Bigl\langle G_{ret}(x'_2,x'_1) \Bigr\rangle$ and hence we have 
$C(x_2,x_1;x'_2,x'_1) \approx 0$. Thus, successive pulses are uncorrelated
and the expected variation in their flight times is given by Eq.~(\ref{eq:Dt}).

    More generally, $C(x_2,x_1;x'_2,x'_1) \not= 0$, and a pair of successive
pulses will be correlated. In this case, the expected variation in their flight 
times is not given by Eq.~(\ref{eq:Dt}), but is expected to be smaller. 
Recall that
the function $\Bigl\langle G_{ret}(x_2,x_1) G_{ret}(x'_2,x'_1) \Bigr\rangle$
is the mean value of the product of the field at $x_2$ due to a delta function
source at $x_1$ with the field at $x'_2$ due to a delta function source at 
$x'_1$. In the absence of metric fluctuations, this quantity is nonzero only
when the propagation times are equal, that is, when $t_2 -t_1 = t'_2 -t'_1$.
In the presence of metric fluctuations, this function is a Gaussian peaked
at the point where  $t_2 -t_1 = t'_2 -t'_1$, and with a width which 
characterizes the expected deviation in propagation times. This behavior will
be illustrated by specific examples in the following section.

\section{Variation in Photon Flight Times}
\label{sec:Flight}

    In this section, we wish to consider the situation illustrated in Fig. 1,
where a photon is emitted at $t=t_1$ and detected at $t=t_2$ and then a second
photon is emitted at $t=t'_1$ and detected at $t=t'_2$. Let $t_0 = t'_1 - t_1$,
the difference in emission times. The effects of metric fluctuations will in
general cause the propagation times, $t_2 -t_1$ and $t'_2 - t'_1$, to differ 
both from the classical propagation time, $r$, and from one another. 
Let $\Delta t$ be the expected deviation from the classical time for either
photon:
\begin{equation}
\Delta t = \sqrt{\langle (t_2 -t_1 - r)^2 \rangle} = 
            \sqrt{\langle (t'_2 -t'_1 - r)^2 \rangle} 
         = \frac{\sqrt{\langle \sigma_1^2 \rangle}}{r} \,.
\end{equation}
Let $\delta t$ be the expected variation in the propagation times of successive
photons:
\begin{equation}
\delta t' = \sqrt{\langle (t_2 -t_1 - t'_2  +t'_1)^2 \rangle} \, .
\end{equation}

\begin{figure}
\epsfysize=10cm\epsffile{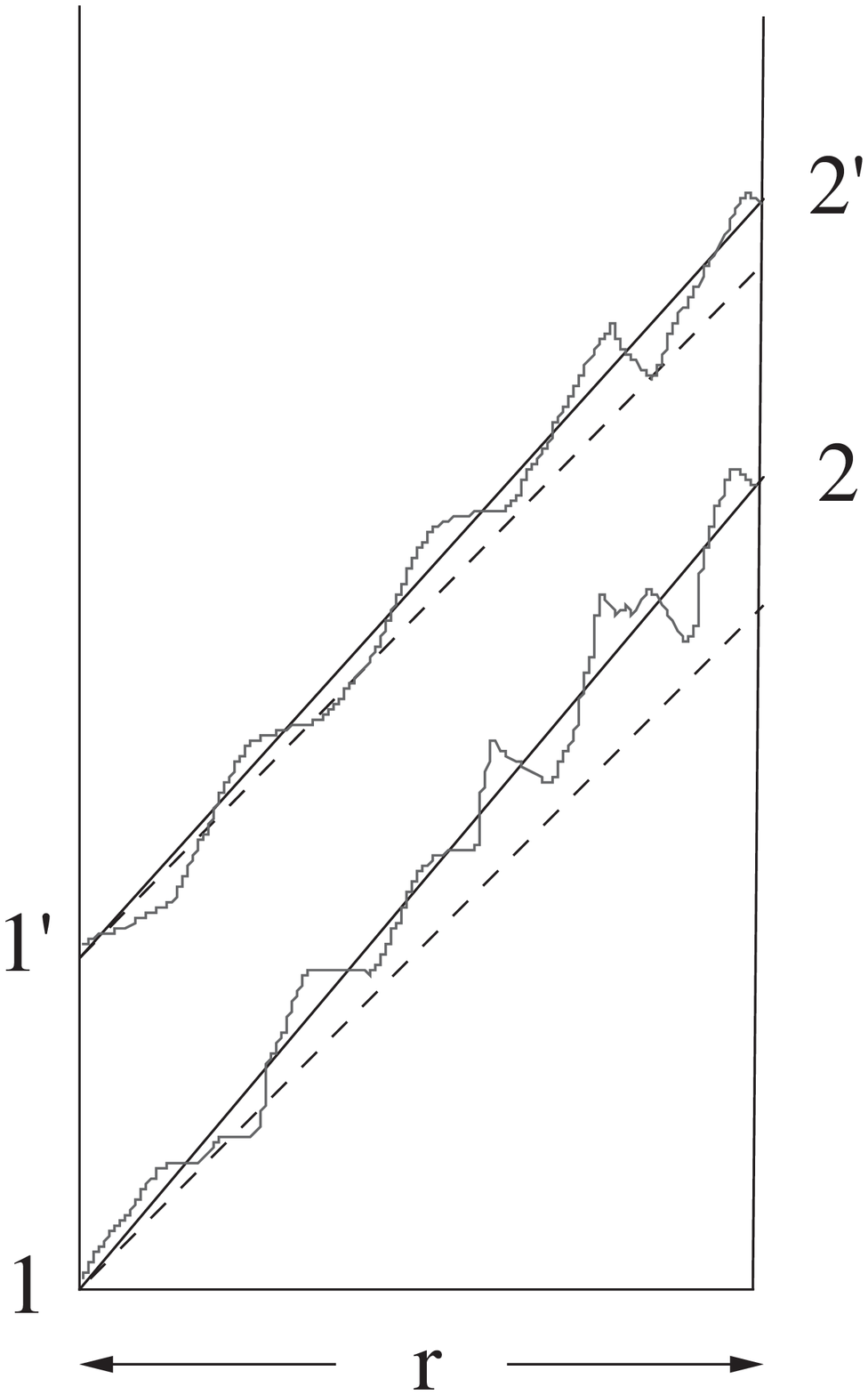}
\begin{caption}[]

 A photon is emitted at point $1$ and detected at point $2$. A second
photon is emitted at point $1'$ and detected at point $2'$. In the absence of 
metric fluctuations, a photon propagates on the classical lightcone, illustrated
by the dashed lines. Metric fluctuations cause the the photon to move on a 
stochastic trajectory with a propagation time which may either larger or 
smaller than the classical flight time. The mean trajectory for a fixed flight
time is illustrated by the solid lines.

\label{Figure-1}  
\end{caption}
\end{figure}\par

When the propagation of the two photons is uncorrelated, $\delta t = \Delta t$.
We expect this to be the case when the difference in emission times, $t_0$, is
sufficiently long. More generally, the photons may be correlated, in which case
$\delta t < \Delta t$. Various special cases will be discussed in the following
subsections. We will need to calculate $\langle \sigma_1 \sigma'_1 \rangle$,
which is given by an expression analogous to Eq.~(\ref{eq:sigma1ex}) : 
\begin{equation}
\langle \sigma_1 \sigma'_1 \rangle =  {1\over {15}}\, r^2 \, 
\int_{0}^{r} dz \int_{0}^{r} dz'
\:\, \langle \varphi(x) \varphi(x') \rangle \, , \label{eq:sigma11'ex}
\end{equation}
where the $z$-integration is taken along the mean path of the first photon,
and the $z'$-integration is taken along that of the second photon. Here
we will assume that $\Delta t \ll r$, so the slopes, $v$ and $v'$, 
of the two mean paths are 
approximately unity. Thus the two-point function in Eq.~(\ref{eq:sigma11'ex})
will be assumed to be evaluated at $\rho = |{\bf x} -{\bf x'}| = |z -z'|$ 
and $\tau = |t - t'| = |z -z' -t_0|$.

\subsection{Large $t_0$}
\label{sec:large_t}

    Let us first consider the limit in which $t_0$ is large compared to either
$r$ or $\lambda_g$, and hence the integration in 
Eq.~(\ref{eq:sigma11'ex}) is over points for which 
$\tau \approx t_0 \gg \rho$ and  $\tau \gg \lambda_g$. In this limit, the  
two-point function becomes approximately [See Eq.~(\ref{eq:grthi_t}).]
\begin{equation}
\langle \varphi(x) \varphi(x') \rangle  \approx 
               \frac{45\, \lambda_g^2\, h^2}{\pi^2 \, \tau^2}\,.
\end{equation}
 Thus, we find
\begin{equation}
\langle \sigma_1 \sigma'_1 \rangle \approx 
             3\, \lambda_g^2\, h^2 \,\frac{(\Delta t)^4}{\pi^2\, t_0^2}
            \, . \label{eq:sigma11'A}
\end{equation}
In this limit, $\langle \sigma_1 \sigma'_1 \rangle \ll 
\langle \sigma_1^2 \rangle$, and we obtain the uncorrelated case. In the long
graviton wavelength (low temperature) limit, $\langle \sigma_1^2 \rangle$ 
is given by Eq.~(\ref{eq:sigma1genlow}) and we have
\begin{equation}
\delta t \approx \Delta t = h\, r\,.
\end{equation}
In the short
graviton wavelength (high temperature) limit, $\langle \sigma_1^2 \rangle$ 
is given by Eq.~(\ref{eq:sigma1gen}) and 
\begin{equation}
\delta t \approx \Delta t = h\, \sqrt{\frac{3}{\pi}\,r \,\lambda_g } \;
\sqrt{\ln \biggl( \frac{r}{\beta} \biggr) + c_1 } \,. \label{eq:dthi}
\end{equation} 

\subsection{Long Graviton Wavelengths}

    In the limit in which $\lambda_g$ is large compared to both $r$ and to
$t_0$, we use the low temperature approximation to the thermal two-point
function, Eq.~(\ref{eq:grtlo}). Substituting this approximate form into 
Eq.~(\ref{eq:sigma1ex2}) yields
\begin{equation}
\langle \sigma_1^2 \rangle \approx  h^2 \, r^4\;
\biggl[1 - \frac{2\, \pi^2\, r^2}{45\, \lambda_g^2} \biggl] 
\, . \label{eq:sigma1low}
\end{equation}
Similarly, we obtain
\begin{equation}
\langle \sigma_1 \sigma'_1 \rangle \approx  \langle \sigma_1^2 \rangle -
 h^2\, \frac{\pi^2\, r^4\, t_0^2}{5\, \lambda_g^2}
\, , \label{eq:sigma11'low}
\end{equation}
and hence combining these two results, we obtain
\begin{equation}
B \approx 2 h^4 \,\frac{\pi^2\, r^8\, t_0^2}{5\,  \lambda_g^6} \,.
\end{equation}

   In the calculation of $B$, it was possible to approximate the slopes 
of the photon trajectories as being unity. In order to obtain nonvanishing
expressions for $\sigma_0$ and $\sigma'_0$, we must consider the deviation of 
these slopes, $v$ and $v'$, from unity. To leading order in 
$|v-1|$ and $|v'-1|$, we have
\begin{eqnarray}
\sigma_0^2 &\approx& (1-v)^2\, r^4 \nonumber \\ 
{\sigma'_0}^2 &\approx& (1-v')^2\, r^4 \nonumber \\ 
\sigma_0 \sigma'_0 &\approx& (1-v)(1-v')\, r^4 \,.  
\end{eqnarray}
For the calculation of $A$, we use only the approximation in which
\begin{equation}
\langle \sigma_1^2 \rangle \approx \langle {\sigma'_1}^2 \rangle 
\approx \langle \sigma_1 \sigma'_1 \rangle \approx h^2\, r^4 \,.
\end{equation}
Then we obtain
\begin{equation}
A \approx h^2\, r^8 (v-v')^2 \, ,
\end{equation}
The magnitude of the argument of the exponential in Eq.~(\ref{eq:Gsq}) becomes
\begin{equation}
C = \frac{A}{2B}  \approx 
                   \frac{5\, \beta^2\, (v-v')^2}{4\, \pi^2\, h^2\, t_0^2} \, .
\end{equation}
 
   In this case, $\langle \sigma_1 \sigma'_1 \rangle$ is not negligible, and  
hence the two photons are correlated.  The mean 
variation in the flight times is given by $\delta t = \delta v\, r$, 
where $\delta v$ is the value of $|v-v'|$ for which $C = 1$. Thus
\begin{equation}
\delta t \approx \frac{2 \pi}{\sqrt{5}} \biggl(\frac{t_0}{\lambda_g}\biggr)
                  h\, r \,. \label{eq:delt}
\end{equation} 
From Eqs.~(\ref{eq:Dt}) and (\ref{eq:sigma1genlow}), we see that
\begin{equation}
\Delta t \approx h\, r \,  \label{eq:Delt}
\end{equation} 
and hence that $\delta t \ll \Delta t$, i.e., the photons are highly correlated.

\subsection{Large Separations}

    In this section, we will consider the case where the separation $r$
is larger than either the graviton wavelength $\lambda_g$ or 
the temporal separation
between the pulses, $t_0$. The integrations in Eq.~(\ref{eq:sigma11'ex}) now
involves pairs of points with timelike, spacelike, and null separations, as
illustrated in Fig. 2. This is contrast to the situation in 
Sect.~\ref{sec:large_t}, where the integrations are entirely over pairs of 
timelike
separated points. We will assume that $r \gg \lambda_g$. Thus we may use the 
high temperature form of the thermal two-point function, which is given by
Eq.~(\ref{eq:grthi_t}) for timelike separations, and Eq.~(\ref{eq:grthi_s}) 
for spacelike separations. Recall
that the former form is valid only for $\tau - \rho > \beta = \lambda_g$. 
Thus the timelike
separated point points will yield a contribution to 
$\langle \sigma_1 \sigma'_1 \rangle$ which is of $O(\ln \lambda_g)$. 
However, the
spacelike separated points yield a contribution which is of 
$O(\lambda_g^{-1})$, and hence is dominant in the small $\lambda_g$ limit. 
Inserting Eq.~(\ref{eq:grthi_s}) into
Eq.~(\ref{eq:sigma11'ex}), and integrating over spacelike separated points 
yields
\begin{equation}
\langle \sigma_1 \sigma'_1 \rangle \approx  \frac{3}{\pi}\, h^2\, \lambda_g\, 
r^2 \int_{0}^{r} dz \int_{0}^{z - t_0} dz'\:\, \frac{1}{\rho} \,
 = \frac{3}{\pi}\, h^2\, \lambda_g\, r^3 \; 
       \biggl[\, \ln\biggl(\frac{r}{t_0}\biggl)\, -\,1 \biggl] \,. 
                                               \label{eq:sigma11'hi}
\end{equation}
Compare this result with Eq.~(\ref{eq:sigma1gen}) for 
$\langle \sigma_1^2 \rangle$ in the $r \gg \lambda_g$ limit. We see that
\begin{equation}
\frac{\langle \sigma_1 \sigma'_1 \rangle^2}{\langle \sigma_1^2 \rangle\,
\langle {\sigma'_1}^2 \rangle} \approx 
\frac{\ln\bigl({r}/{t_0}\bigl)}
                     {\ln\bigl({r}/{\lambda_g}\bigl)} \, .
                                               \label{eq:ratio}
\end{equation}
Thus far, we have not specified the relative magnitudes of 
$\lambda_g$ and $t_0$.
Let first consider the case $\lambda_g \ll t_0$. In this limit, the ratio in
Eq.~(\ref{eq:ratio}) slowly approaches zero. Thus, in this case, the two
photons become uncorrelated. Because of the slow rate at which this ratio
vanishes, we might describe this case as one in which the events are weakly
uncorrelated, and the variation in the flight times is approximately given
by Eq.~(\ref{eq:dthi}).

\begin{figure}
\epsfysize=10cm\epsffile{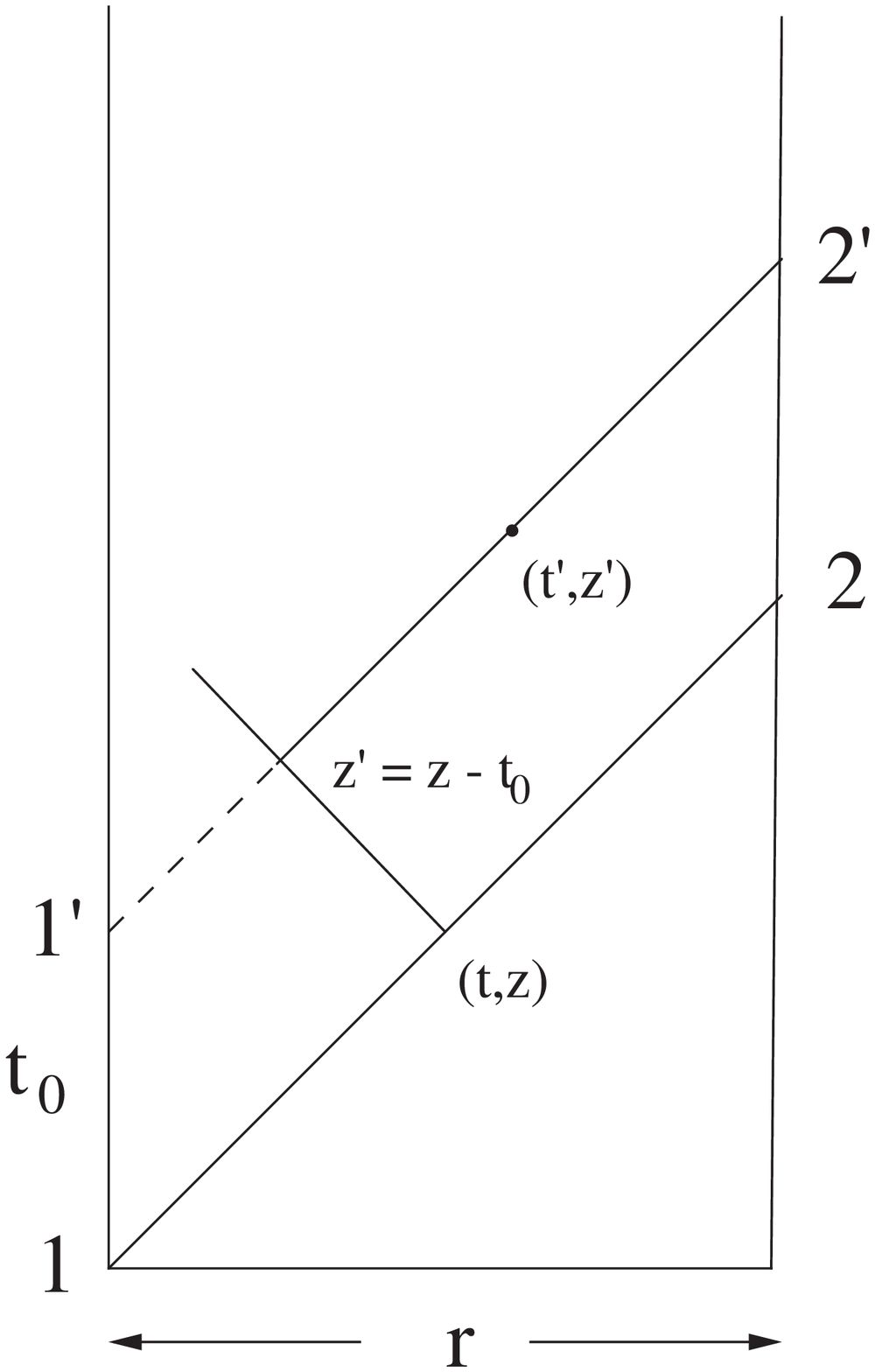}
\begin{caption}[]
  
 The domain of integration in Eq.~(\ref{eq:sigma11'ex}) is illustrated 
for the case that $r > t_0$. For a given point $(t,z)$ on the first
trajectory, a point $(t',z')$ on the second trajectory may spacelike separated
(dashed line) or timelike separated (solid line) from the first point. The
boundary case of null separation occurs at $z' = z - t_0$. In the high 
temperature limit, $r \gg \beta$, the spacelike portion yields the dominant
contribution to the integral.

\label{Figure-2}  
\end{caption}
\end{figure}\par

The remaining possibility is the case where $\lambda_g > t_0$. The ratio in
Eq.~(\ref{eq:ratio}) is now larger than unity, and hence $B < 0$. In this
case, the derivation of the correlation function given in 
Sect.~\ref{sec:corrfnt} breaks down, and our results are inconclusive.

\section{Effects of Cosmological Gravitons}
\label{sec:Bounds}

   In principle, the results of the previous sections could be used to search
for relict cosmological gravitons. It is expected that there may exist a
bath of gravitons at the present time which was created in the early universe.
For example, if the universe evolved from a state of thermal equilibrium at the
Planck epoch without inflation, one would expect that there should now be a 
thermal bath of gravitons at a temperature of approximately $3K$. On the other
hand, inflation would tend to wipe out such a bath, but could create more
gravitons at the end of inflation. Inflation at an energy scale of the order
of $10^{15}$ GeV might produce a bath with $\lambda_g \approx 10^4 {\rm cm}$
and $h \approx 10^{-36}$ \cite{F95,F87}. The effects of either of these baths
upon the propagation time of photons is far too small to be detectable,
however. 

We can take a different approach in which we conjecture that some unknown
process may have generated a much larger bath of gravitons, and seek 
observational bounds on such a bath. The timing data from pulsars provides
one such set of bounds. This data has in fact been used in recent years to
place limits on a background of classical gravity waves 
\cite{Sazhin}-\cite{Taylor}. Thus, in the next subsection, we will briefly
review the effect of a classical gravity wave upon the observed arrival
times of pulses. 

\subsection{Classical Gravity Waves and Photon Flight Times}
\label{sec:classical}

  Consider a classical metric perturbation, $h_{ij}$, in the 
transverse-tracefree (TT) gauge. If one photon is emitted at $t=0$ and a second
at $t=t_0$, the difference in their flight times over a spatial distance $r$
is (See Ref. \cite{Sazhin} and Eq.~(46) in I.) 
\begin{equation}
\Delta \, T = -\frac{1}{2} \int_0^r [H(t_0+z,z) \, -\, H(z,z)]\, dz \,,
\end{equation}
where $H = H(t,z) = h_{ij} n^i n^j$, and $n^i$ is the unit three-vector in the 
direction of the photons' propagation. Let us assume that both the source and 
the detector are initially at rest with respect to our coordinate system, so 
their four-velocities are $u^\mu = \delta_t^\mu$. From the geodesic equation
\begin{equation}
\frac{d u^\mu}{d\, \tau} = - \Gamma^\mu_{\alpha \beta}\, u^\alpha\, u^\beta
                         = 0 \,.
\end{equation}
The second step follows from the fact that $\Gamma^\mu_{tt} = 0$ in the TT
gauge. Thus to linear order, the metric perturbation does not change the
four-velocities of either the source or the detector, and hence $\Delta \, T$ 
is not only a coordinate time difference, but also the proper time difference
in photon arrival times due to the gravity wave.

In order to give simple estimates of the magnitude of $\Delta \, T$ in various 
limits, let us take an explicit form for $H$:
\begin{equation}
H(t,z) = h\, \cos(k_z z - \omega t + \delta) \, .
\end{equation}
[Note that the amplitude, $h$, of the classical wave may differ from the
amplitude defined in Eq.~(\ref{eq:h2def}) by small numerical factors which 
we will ignore.]
Here we are assuming that the photons propagate along the line $x = y =0$, and
that the gravity wave travels in some other direction, so that $k_z \not= 
\omega$. Our result for $\Delta \, T$ now becomes
\begin{equation}
\Delta \, T = \frac{h}{2(\omega - k_z)} \Bigl \{ 
\sin\bigl[(k_z -\omega)r -\omega t_0 +\delta \bigl] - 
\sin( \delta -\omega t_0) -\sin(k_z -\omega)r +\delta) + \sin\delta \Bigr\}
 \, . \label{eq:Del_T}
\end{equation}
If neither $t_0$ nor $r$ is small compared to $\lambda_g$, the quantity
within braces in the above expression is of order unity, so
\begin{equation}
\frac{\Delta \, T}{t_0} \approx h\,\frac{\lambda_g}{t_0}\,. \label{eq:Del_T1}
\end{equation}
If $t_0 \ll \lambda_g$ in Eq.~(\ref{eq:Del_T}), then
\begin{equation}
\Delta \, T \approx \frac{h\,t_0\,\omega}{2(\omega - k_z)} \Bigl \{ 
\cos\delta - \cos\bigl[(k_z -\omega)r +\delta \bigl] \Bigr\}
 \, , \label{eq:Del_T2}
\end{equation}
and hence
\begin{equation}
\frac{\Delta \, T}{t_0} \approx h \,.
\end{equation}
Similarly, if $r \ll \lambda_g$ in Eq.~(\ref{eq:Del_T}), then
\begin{equation}
\Delta \, T \approx \frac{1}{2}h\, r \Bigl [ 
\cos\delta - \cos(\delta - \omega t_0) \Bigr]
 \, , \label{eq:Del_T3}
\end{equation}
and hence
\begin{equation}
\frac{\Delta \, T}{t_0} \approx h \frac{r}{t_0}\,.
\end{equation}
Finally, if $t_0 \ll \lambda_g$ and $r \ll \lambda_g$, then 
Eq.~(\ref{eq:Del_T}) becomes
\begin{equation}
\Delta \, T \approx -\frac{1}{2}h\, r\, t_0\, \sin\delta \, , 
 \label{eq:Del_T4}
\end{equation}
and 
\begin{equation}
\frac{\Delta \, T}{t_0} \approx h \frac{r}{\lambda_g}\,.
\end{equation}
Our results for the various cases, as well as the results of previous
sections for $\Delta t$ and $\delta t$, are summarized in Table 1. 

\begin{table}
\begin{tabular}{|c||c|c|c|c|} \cline{1-4}
 & $\frac{\Delta T}{t_0}$ & $\frac{\Delta t}{t_0}$ & $\frac{\delta t}{t_0}$ \\
\hline
$\lambda_g \gg r \gg t_0$ & $h\,\frac{r}{\lambda_g}$ & & 
      $h\, \frac{r}{\lambda_g}$ & correlated \\
\cline{1-1} 
$\lambda_g \gg t_0 \gg r$ & & $h\, \frac{r}{t_0}$ & &  \\
\cline{1-2} \cline{4-5} 
$t_0 \gg\lambda_g \gg r$ &$h\, \frac{r}{t_0}$ & & $h\, \frac{r}{t_0}$ & \\
\cline{1-4}
$t_0 \gg r \gg\lambda_g$ & $h\, \frac{\lambda_g}{t_0}$ & &
            $h\, \frac{\sqrt{r\,\lambda_g}}{t_0}$ & uncorrelated \\
\cline{1-1}
$r \gg t_0 \gg \lambda_g$ & & $h\, \frac{\sqrt{r\,\lambda_g}}{t_0}$ & &  \\
\cline{1-2} \cline{4-5}
$r \gg \lambda_g \gg t_0$ & $h$ & & ?? \\
\cline{1-4}
\end{tabular}
   \caption{Summary of results for $\Delta T$, $\Delta t$, and $\delta t$.}
\end{table}

Note that $\Delta T$ and $\delta t$ are approximately equal in the limit
of long wavelengths, $\lambda_g \gg r$. As a matter of principle, the
two quantities are quite different: $\delta t$ arises from quantum fluctuations
of the lightcone, whereas $\Delta T$ is calculated in a fixed classical
spacetime with a precisely defined lightcone. This is reflected in the fact 
that when $\lambda_g \ll r$, the short wavelength limit, $\delta t$ grows
with increasing $r$, whereas $\Delta T$ does not.

\subsection{Bounds on Graviton Baths}

Pulsars can be exceptionally stable astronomical clocks. Timing data from
a number of stable pulsars have been gathered over the past decade 
\cite{Strinebring,Taylor} which
places an upper bound on both $\Delta T/t_0$ and $\delta t/t_0$ of the order
of $10^{-14}$ with $t_0 \approx 10\,{\rm yrs}$. This data may be used to
place limits upon the present-day energy density in both classical gravity 
waves and in non-classical gravitons. This energy density is
\begin{equation}
\rho_g = \frac{1}{2}\, \langle h_{{ij}_{,t}}\, h^{ij}_{,t} \rangle
       \approx 60\, \pi^2\, \frac{h^2}{\lambda_g^2} \,,
\end{equation}
where we have used Eq.~(\ref{eq:h2def}) in the second step. This relation may 
be written as
\begin{equation}
h \approx 1.5 \times 10^{-27} \, \biggl(\frac{\lambda_g}{1 {\rm cm}} \biggr) \, 
    \sqrt{\frac{\rho_g}{10^{-30} {\rm g/cm}^3}}  \, . \label{eq:hrho}
\end{equation}

Typical pulsar distances are of order $r \approx 1\, {\rm kpc} \approx 3 \times 
10^{21} {\rm cm}$. If we consider the case $\lambda_g < 10\,{\rm yrs} \approx
10^{19} {\rm cm}$, then we have that $r \gg t_0 \gg \lambda_g$. In this case, 
from Eq.~(\ref{eq:Del_T1}), we obtain a bound on the energy density in classical
gravity waves of 
\begin{equation}
\rho_g < 5 \times 10^{-43} {\rm \frac{g}{cm^3}} \, 
           \biggl( \frac{10^{19} {\rm cm}}{\lambda_g} \biggr)^4 \,.
\end{equation}
If, for example, $\lambda_g = 0.14 \,{\rm yrs}$, we have $\rho_g < 
10^{-36} {\rm {g}/{cm^3}}$, in agreement with the more careful analysis of
Strinebring, {\it et al} \cite{Strinebring}.

The corresponding bound upon gravitons in a squeezed vacuum state is somewhat
more stringent. For the case $r \gg t_0 \gg \lambda_g$,
\begin{equation}
\frac{\delta t}{t_0} \approx 
                   \sqrt{\frac{r}{\lambda_g}}\; \frac{\Delta T}{t_0} \,.
\end{equation}
For gravitons, we obtain
\begin{equation}
\rho_g <  10^{-45} {\rm \frac{g}{cm^3}} \, 
           \biggl( \frac{10^{19} {\rm cm}}{\lambda_g} \biggr)^3 \,.
\end{equation}

In the long wavelength case, $\lambda_g \gg r$, we have $\Delta T/t_0
\approx \delta t/t_0 \approx h r/\lambda_g$. The resulting bound on
both classical gravity waves and gravitons is 
\begin{equation}
\rho_g < 4 \times 10^{-48} {\rm \frac{g}{cm^3}} \,,
\end{equation}
and is independent of $\lambda_g$, as long as $\lambda_g \gg 1\, {\rm kpc}$.

\section{Summary and Discussion}
\label{sec:Summ}

In this paper, we have obtained a result for the mean squared fluctuations
of the geodesic interval function, $\langle \sigma_1^2 \rangle$, in the case
of a thermal bath at high temperature. This result can also be used to discuss 
graviton baths with an approximately Planckian spectrum, but with an arbitrary
amplitude. We have found a general expression for the correlation function,
Eq.~(\ref{eq:Gsq}). This correlation function was used to determine when a
pair of photon trajectories is uncorrelated, so that the mean variation
in successive photon flight times, $\delta t$, is essentially equal to 
$\Delta t$, the mean deviation from the classical flight time.
When the trajectories are in fact correlated, this function also enables us
to compute  expressions for $\delta t$. Application of these result to pulsar
timing data seems to lead to some nontrivial bounds upon the allowed amplitudes
of baths of long wavelength gravitons. Although the baths of gravitons in a
squeezed vacuum state that one might expect to have been created in the early 
universe are unobservable by this means, these arguments limit such gravitons
created by unknown mechanisms. This presumably places some restrictions upon 
the initial quantum state of the universe.

\vspace {0.5 in}
{\bf Acknowledgement:} We would like to thank Alex Vilenkin for helpful
comments. This work was supported in part by the National
Science Foundation under Grant PHY-9507351 and by Conselho Nacional de
Desevolvimento Cientifico e Tecnol{\'o}gico do Brasil (CNPq).

\appendix
\section*{Appendix}
\setcounter{equation}{0}
\renewcommand{\theequation}{A\arabic{equation}}

   Here we summarize some of the properties of the thermal two-point 
function in coordinate space. The renormalized thermal two-point function,
i.e. the finite temperature function minus the vacuum contribution, is equal
to the Hadamard function, as discussed in Sec.~\ref{sec:Bath}, and is
expressible as an image sum:
\begin{equation}
G_{RT}(x,x') = \frac{1}{4 \pi^2}\: {\sum_{n = -\infty}^{\infty}}'\,
               \frac{1}{\rho^2 - (\tau +i n \beta)^2} \, ,
\end{equation}
where $\rho = |{\bf x} - {\bf x}'|$ and $\tau = |t -t'|$, and the prime
on the summation indicates that the $n = 0$ term is omitted. This sum
may be evaluated by means of the Poisson summation formula, which states
that
\begin{equation}
{\sum_{n = -\infty}^{\infty}} f(n) = {\sum_{n = -\infty}^{\infty}} {\hat f}(n)
 \,,
\end{equation}
where ${\hat f}(n) = \int_{-\infty}^{\infty} {\rm e}^{-2\pi i n x} \,f(x)\,
{\rm d} x$ is the Fourier transform of $f$. 

In our case, $f(x) = (4 \pi^2)^{-1} [\rho^2 - (\tau +i n \beta)^2]^{-1}$. 
If $\tau > \rho$, then $\hat f(n) = 0$ for $n \geq 0$, and 
\begin{equation}
\hat f(n) = \frac{1}{4\pi \rho \beta}\, 
\biggl[{\rm e}^{2\pi(\tau +\rho)n/\beta} \;-\; 
                      {\rm e}^{2\pi(\tau +\rho)n/\beta} \biggl] \,,
\end{equation}
for $n < 0$. Similarly, if $\tau < \rho$, then 
\begin{equation}
\hat f(n) = \frac{1}{4\pi \rho \beta}\, 
{\rm e}^{2\pi(\tau -\rho)n/\beta} \,,
\end{equation}
for $n \geq 0$, and
\begin{equation}
\hat f(n) = \frac{1}{4\pi \rho \beta}\, 
{\rm e}^{2\pi(\tau +\rho)n/\beta} \,,
\end{equation}
for $n < 0$. We may now evaluate $G_{RT}$. In both the
$\tau > \rho$ and $\tau < \rho$ cases, we find the same result:
\begin{equation}
G_{RT}(x,x') = \frac{1}{4 \pi^2}\: \Biggl\{ \frac{\pi}{\beta \,\rho}\: \biggl[
\frac{1}{{\rm e}^{2\pi(\tau +\rho)/\beta} - 1} \;-\; 
\frac{1}{{\rm e}^{2\pi(\tau -\rho)/\beta} - 1} \biggl] \; + \;
    \frac{1}{\tau^2 - \rho^2} \Biggl\} \,. \label{eq:grt}
\end{equation}
 
In the low temperature limit, $\beta \rightarrow \infty$, $G_{RT}$ has the
asymptotic form
\begin{equation}
G_{RT}(x,x') \sim \frac{1}{12 \beta^2}\: \biggl(1\, -\, 
                   \pi^2 \, \frac{\rho^2 + 3\tau^2}{15 \beta^2} \biggr) + 
                     O(\beta^{-6}) \,. \label{eq:grtlo}
\end{equation}
In the high temperature limit, $\beta \rightarrow 0$, for non-null separated
points ($\rho \not= \tau$), $G_{RT}$ has the asymptotic form
\begin{equation}
G_{RT}(x,x') \sim \frac{1}{4 \pi^2}\: \Biggl\{ - 
\frac{\pi}{\beta \,\rho \Bigl[{\rm e}^{2\pi(\tau -\rho)/\beta} - 1 \Bigr] }
\; + \;
    \frac{1}{\tau^2 - \rho^2} \Biggl\} \,. \label{eq:grthi}
\end{equation}
For timelike separations, ($\tau > \rho$), this yields
\begin{equation}
G_{RT}(x,x') \sim \frac{1}{4 \pi^2 (\tau^2 - \rho^2)}  \,. \label{eq:grthi_t}
\end{equation}
Similarly, for spacelike separations, ($\tau < \rho$), we have
\begin{equation}
G_{RT}(x,x') \sim \frac{1}{4 \pi \beta \rho} + O(\beta^0) \,. \label{eq:grthi_s}
\end{equation}
 The form of $G_{RT}$ on the lightcone is obtained from Eq.~(\ref{eq:grt})
by taking the limit $\tau \rightarrow \rho$:
\begin{equation}
G_{RT}(x,x') \rightarrow \frac{1}{4 \pi^2}\: 
\Biggl\{ \frac{\pi}{\beta \,\rho}\: \biggl[
\frac{1}{{\rm e}^{4\pi \rho/\beta} - 1} \;+\; 
\frac{1}{2} \biggl] \; - \;  \frac{1}{\rho^2} \Biggl\} \,. \label{eq:grt_lc}
\end{equation}
As required, $G_{RT}$ is finite on the lightcone. Let us now take the high
temperature limit of this lightcone form to obtain
\begin{equation}
G_{RT}(x,x') \sim \frac{1}{8 \pi \beta \rho} + O(\beta^0) \,. \label{eq:grthi_n}
\end{equation}
Comparison of Eqs.~(\ref{eq:grthi_s}) and ~(\ref{eq:grthi_n}) reveals that 
the high temperature limit of $G_{RT}$ is discontinuous on the lightcone.


\begin{thebibliography}{--}

\bibitem{F95} L.H. Ford, Phys. Rev. D {\bf 51}, 1692 (1995).

\bibitem{pulse} In this paper we will ignore any effects due to the finite
sizes of photon wavepackets.

\bibitem{units} Units in which $\hbar = c = 16 \pi G =1$ will be used in this
paper. The metric signature will be $(1,-1,-1,-1)$.

\bibitem{Grishchuk} L.P. Grishchuk and Y.V. Sidorov, Phys. Rev. D {\bf 42},
3413 (1990).

\bibitem{F82} L.H. Ford, Ann. Phys. (NY) {\bf 144}, 238 (1982).

\bibitem{Kuo} C.-I Kuo and L.H. Ford, Phys. Rev. D {\bf 47}, 4510 (1993).

\bibitem{FP} L.H. Ford and L. Parker, Phys. Rev. D {\bf 16}, 1601 (1977).

\bibitem{F87} L.H. Ford, Phys. Rev. D {\bf 35}, 2955 (1987).

\bibitem{Sazhin} M.V. Sazhin, Astron. Zh. {\bf 55}, 65 (1978) [Sov. Astron.
{\bf 22}, 36 (1978)].

\bibitem{Det} S. Detweiler, Ap. J. {\bf 234}, 1100 (1979).

\bibitem{Strinebring} D.R. Strinebring, M.F. Ryba, J.H. Taylor, and R.W. Romani,
Phys. Rev. Lett. {\bf 65}, 285 (1990).

\bibitem{Taylor} J.H. Taylor, Phil. Trans. R. Soc. Lond. A {\bf 341}, 
117 (1992).

\end{thebibliography}
\end{document}